\documentclass[aps,twocolumn,showpacs,superscriptaddress,nofootinbib,showkeys]{revtex4}
\usepackage{graphicx}
\usepackage{dcolumn}
\usepackage{html}

\newcommand{\uno}{{\bf 1}}
\newcommand{\dos}{{\bf 2}}
\newcommand{\tres}{{\bf 3}}
\newcommand{\cuatro}{{\bf 4}}
\newcommand{\seis}{{\bf 6}}
\newcommand{\ocho}{{\bf 8}}
\newcommand{\diez}{{\bf 10}}
\newcommand{\veinte}{{\bf 20}}
\newcommand{\veintisiete}{{\bf 27}}
\newcommand{\treintaycinco}{{\bf 35}}
\newcommand{\cincuentayseis}{{\bf 56}}
\newcommand{\setenta}{{\bf 70}}
\newcommand{\cientoochentaynueve}{{\bf 189}}
\newcommand{\doscientosochenta}{{\bf 280}}
\newcommand{\cuatrocientoscinco}{{\bf 405}}
\newcommand{\setecientos}{{\bf 700}}
\newcommand{\milcientotreintaycuatro}{{\bf 1134}}
\newcommand{\dosmilseiscientosnoveintaycinco}{{\bf 2695}}

\def\slashchar#1{\setbox0=\hbox{$#1$}
   \dimen0=\wd0 \setbox1=\hbox{/} \dimen1=\wd1
   \ifdim\dimen0>\dimen1 \rlap{\hbox to \dimen0{\hfil/\hfil}} #1
   \else  \rlap{\hbox to \dimen1{\hfil$#1$\hfil}} / \fi}
\begin{document}
\title{SU(6) Extension of the Weinberg-Tomozawa Meson-Baryon Lagrangian}
\author{C. Garc{\'\i}a-Recio}
\email{g_recio@ugr.es}
\author{J. Nieves}
\email{jmnieves@ugr.es}
\author{L.L. Salcedo}
\email{salcedo@ugr.es}
\affiliation{
Departamento de F{\'\i}sica At\'omica, Molecular y Nuclear,
Universidad de Granada, E-18071 Granada, Spain
}
\begin{abstract}
A consistent SU(6) extension of the Weinberg-Tomozawa meson-baryon
chiral Lagrangian is constructed which incorporates vector meson
and baryon decuplet
degrees of freedom. The corresponding Bethe-Salpeter approximation
predicts the existence of an isoscalar spin-parity $\frac{3}{2}^-$
$K^*N$ bound state (strangeness +1) with a mass around
1.7--1.8$\,$GeV. It is the highest hypercharge state of an
antidecuplet SU(3) representation and it is unstable through $K^*$
decay. The estimated width of this state (neglecting $d$-wave $KN$
decay) turns out to be small ($\Gamma \le 15\,$MeV). Clear signals of
this resonance would be found in reactions like $\gamma p \to {\bar
K}^0 p K^+ \pi^-$ by looking at the three body $p K^+ \pi^-$ invariant
mass.
\end{abstract}



\date{\today}
\maketitle



\section{Introduction}

Forty years ago, it was
suggested~\cite{Gursey:1992dc,Pais:1964,Sakita:1964qq} that it might
be a useful approximation to assume that the light quark--light quark
interaction is approximately spin independent as well as SU(3)
independent. This corresponds to treating the six states of a light
quark ($u$, $d$ or $s$ with spin up, $\uparrow$, or down,
$\downarrow$) as equivalent, and leads us to the invariance group
SU(6). In order that we can speak meaningfully of SU(6)
transformations affecting spin but not orbital angular momentum ($L$)
as invariances, it must be assumed that the orbital angular momentum
and the quark spin are to a good approximation, separately
conserved. This, in turn requires the spin--orbit, tensor and
spin--spin interactions between quarks to be small. As is known,
mixing the compact, purely internal flavor symmetry, with the
noncompact Poincare symmetry of spin angular momentum led to some
inconsistencies, which gave rise to the no--go Coleman--Mandula
theorem~\cite{Coleman:1967ad} forbidding such exact hybrid symmetries,
unless supersymmetry is invoked. However, there exist several SU(6)
predictions (relative closeness of baryon octet and decuplet masses,
the axial current coefficient ratio $F/D=2/3$, the magnetic moment
ratio $\mu_p/\mu_n=-3/2$) which are remarkably well satisfied in
nature~\cite{Lebed:1994ga}. This suggests that SU(6) could be a good
approximate symmetry. Indeed, in the large $N_c$ limit (being $N_c$
the number of colors)~\cite{'tHooft:1973jz,Witten:1979kh}, there
exists an exact spin--flavor symmetry for ground state baryons
\cite{Dashen:1993jt}. Moreover, though in general the spin--flavor
symmetry is not exact for excited baryons even in the large $N_c$
limit, in the real world ($N_c=3$), the zeroth order spin--flavor
symmetry breaking turns out to be similar in magnitude to ${\cal O}
(N_c^{-1})$ breaking effects~\cite{Goity:2002pu}. In the meson sector,
an underlying static chiral U(6)$\times$U(6) symmetry has been
advocated by Caldi and Pagels \cite{Caldi:1975tx,Caldi:1976gz}, in
which vector mesons would be ``dormant'' Goldstone bosons acquiring
mass through relativistic corrections. This scheme solves a number of
theoretical problems in the classification of mesons and also makes
predictions which are in remarkable agreement with the experiment. In
any case, although spin-flavor symmetry in the meson sector is not a
direct consequence of large $N_c$ QCD, vector mesons ($K^*,
\rho,\omega, {\bar K}^{*}, \phi$) do exist, they will couple to
baryons and presumably will influence the properties of the baryonic
resonances.  Lacking better theoretically founded models to include
vector mesons, we regard the spin-flavor symmetric scenario as
reasonable first step. The large $N_c$ consequences of this scheme
have been pursued in \cite{Garcia-Recio:2006wb}.

Since the pure SU(3) (flavor) transformations commute with the pure
SU(2) (spin) transformations, it follows that a SU(6) multiplet can be
decomposed into SU(3) multiplets each of definite total spin. With the
inclusion of spin there are 36 quark--antiquark ($q\bar{q}$) states,
and the SU(6) group representation reduction (denoting the SU(6)
multiplets by their dimensionality and a SU(3) multiplet $\mu$ of spin
$J$ by $\mu_{2J+1}$) reads
\begin{equation}
\seis\otimes \seis^* = \treintaycinco \oplus\uno = 
(\ocho_\uno \oplus \ocho_\tres \oplus \uno_\tres) \oplus \uno_\uno \,.
\end{equation}
We might expect the lowest bound state to be a $s$-state and since the
relative parity of a fermion--antifermion pair is odd, the octet of
pseudoscalar ($K, \pi,\eta, {\bar K}$) and the nonet of vector ($K^*,
\rho,\omega, {\bar K}^{*}, \phi$) mesons are commonly placed in the
$\treintaycinco$ representation. Note that the $\treintaycinco$ allows
nine vector mesons but only eight $0^-$ mesons. A ninth $0^-$ meson
($\eta^\prime$) must go in the $\uno$ of SU(6). This may account for the
phenomenological evidence that the mixing of the octet and singlet
states is much smaller for the $0^-$ mesons than for the $1^-$ mesons
\cite{Eidelman:2004wy}. Mesons of spin greater than one can be
understood as states of the $q{\bar q}$ system with $L>0$.  On the
other hand, with the inclusion of the spin there are 216 three quark
states, and it follows
\begin{eqnarray}
&& \seis\otimes \seis \otimes \seis = \cincuentayseis \oplus \veinte 
\oplus \setenta \oplus \setenta  =
 \\
&& (\ocho_\dos \oplus \diez_\cuatro) \oplus (\uno_\cuatro\oplus \ocho_\dos)
\oplus 2\times (\diez_\dos \oplus \ocho_\cuatro\oplus \ocho_\dos \oplus \uno_\dos)  \,.
\nonumber
\end{eqnarray}
It is natural to assign the lowest--lying baryons to the
$\cincuentayseis$ of SU(6), since it can accommodate an octet of
spin--$1/2$ baryons and a decuplet of spin--$3/2$ baryons, which are
precisely the SU(3)--spin combinations of the low--lying baryon states
($(N,\Sigma,\Lambda, \Xi)$ and ($\Delta$, $\Sigma^*$, $\Xi^*$,
$\Omega$)). Furthermore, the $\cincuentayseis$ of SU(6) is totally
symmetric, which allows the baryon to be made of three quarks in
$s$-wave (the color wavefunction being antisymmetric).

Here we will consider describing the $s$-wave interaction between the
lowest--lying meson multiplet ($\treintaycinco$) and the lowest--lying baryons
($\cincuentayseis$-plet) at low energies. At larger energies higher
partial waves are involved and a suitable treatment of spin-orbit
effects in the SU(6) scheme would be required. Thus, assuming that the
$s$-wave effective meson--baryon Hamiltonian is SU(6) invariant, and
since the SU(6) decomposition of the product of the $\treintaycinco$
(meson) and $\cincuentayseis$ (baryon) representations yields
\begin{eqnarray}
\treintaycinco \otimes \cincuentayseis = \cincuentayseis \oplus 
\setenta \oplus \setecientos \oplus \milcientotreintaycuatro,
\end{eqnarray}
we have only four (Wigner-Eckart irreducible matrix elements) free
functions of the meson--baryon Mandelstam variable $s$. Similar ideas
were already explored in the late sixties, within the effective range
approximation~\cite{Carey:1968}. Here, we introduce two major improvements:
i) We make use of the underlying Chiral Symmetry (CS), which allows us
to determine the value of the SU(6) irreducible matrix elements from
the Weinberg-Tomozawa (WT) interaction~\cite{Weinberg:1966kf,Tomozawa:1966jm}, the leading
term of the chiral Lagrangian involving Goldstone bosons and the octet
of spin--$1/2$ baryons.  This is not a trivial fact and it is
intimately linked to the underlying group structure of the WT
term. ii) We go beyond the effective range approximation and follow a
scheme based on the solution of the Bethe-Salpeter-Equation (BSE),
which incorporates two-body coupled channel unitarity and has been
successfully employed in the study of $s$-wave 
($\ocho_\uno$)meson--($\ocho_\uno$)meson and 
($\ocho_\uno$)meson--($\ocho_\dos$, $\diez_\cuatro$)baryon scattering and
resonances, within different renormalization
schemes~\cite{Kaiser:1995eg,Nieves:1998hp,Nieves:1999bx,Oset:1997it,Lutz:2003fm,Jido:2003cb,%
Nieves:2001wt,Garcia-Recio:2002td,Garcia-Recio:2003ks,Kolomeitsev:2003kt,Sarkar:2004jh}.

\section{SU(6) Meson--Baryon Effective Interaction Matrix}

We will work with well defined total isospin ($I$), angular momentum
($J$) and hypercharge ($Y$) (strangeness plus baryon numbers)
meson--baryon states constructed out of the SU(6) $\treintaycinco$
(mesons) and $\cincuentayseis$ (baryon) multiplets. In what follows we
always use the labels $\mu$ and $\phi$ to denote generic SU(3) and
SU(6) representations, respectively. For short, we use the notation
${\cal M}\equiv \left [(\mu_M)_{2J_M+1}, I_M, Y_M\right]$ for mesons
and similarly for baryons (${\cal B}$). Thus, $\mu_{M}=\ocho,\uno$ and
$\mu_{B}=\ocho,\diez$ are the meson and baryon SU(3) multiplets, respectively
and $J_{M,B},I_{M,B},Y_{M,B}$ are the spin, isospin and hypercharge
quantum numbers of the involved hadrons. The meson-baryon states in
terms of the SU(6) coupled (orthonormal) basis read
\begin{eqnarray}
|{\cal M} {\cal B}; JIY \rangle= \sum_
{\mu,\alpha,\phi} 
\left( \begin{array}{cc|c} \mu_M& \mu_B&
   \mu \\ I_MY_M & I_B Y_B &
   IY\end{array}\right) 
 \nonumber \\ 
\times
\left( \begin{array}{cc|c} \treintaycinco& \cincuentayseis&
   \phi \\ \mu_M J_M & \mu_B J_B &
   \mu J\alpha \end{array} \right)
  \left| \phi; \mu_{2J+1}^\alpha IY \right\rangle 
\,,
  \label{eq:estado}
\end{eqnarray}
where $Y=Y_M+Y_B$, $|I_M-I_B| \le I \le I_M+I_B$, and for $s$-wave
scattering $|J_M-J_B| \le J \le J_M+J_B$, while
$\phi=\cincuentayseis,\setenta,\setecientos,\milcientotreintaycuatro$,
and $\alpha$ accounts for the multiplicity of each of the $\mu_{2J+1}$
SU(3) multiplets of spin $J$ (for $L=0$, $J$ is given by the total
spin of the meson--baryon system) entering in the representation
$\phi$. Multiplicities higher than one only happen for the
$\milcientotreintaycuatro$ representation, where the
$\veintisiete_\cuatro$, $\diez_\cuatro$, $\veintisiete_\dos$ and the
$\diez_\dos$ multiplets appear twice and the $\ocho_\cuatro$ and
$\ocho_\dos$ ones appear three times. The index $\mu$ runs over the
($\veintisiete$, $\diez$, $\diez^*$, $\ocho_s$, $\ocho_a$, $\uno$),
($\treintaycinco$, $\veintisiete$, $\diez$, $\ocho$), ($\ocho$) and 
($\diez$) SU(3)
representations for the octet--octet, octet--decuplet, singlet--octet
and singlet--decuplet decompositions, respectively. Finally in
Eq.~(\ref{eq:estado}), the two coefficients which multiply each
element of the SU(6) coupled basis are the SU(3) isoscalar
factors~\cite{deSwart:1963gc}, and the SU(6)--multiplet coupling
factors~\cite{Carter:1969}. The assumption that the effective $s$-wave
meson--baryon {\it potential} ($V$) is a SU(6) invariant operator
implies that i) the coupled states $\left| \phi; \mu_{2J+1}^\alpha IY
\right\rangle$ are eigenvectors of $V$ and ii) the corresponding
eigenvalues $V_\phi(s)$ may depend on the SU(6) representation $\phi$
but not on the other quantum numbers $\mu$, $\alpha$, $J$, $I$, or
$Y$. Thus, in the non-coupled basis we find
\begin{eqnarray}
\langle {\cal M}^\prime  {\cal B}^\prime ; JIY | V
| {\cal M} {\cal B} ; JIY \rangle &=&
\sum_{\phi} V_{\phi}(s) {\cal P}_{{\cal M}{\cal B}, {\cal M}^\prime 
{\cal B}^\prime }^{\phi,JIY} ,
\label{eq:su6}
\end{eqnarray}
where
\begin{eqnarray}
&&
{\cal P}_{{\cal M}{\cal B}, {\cal M}^\prime 
{\cal B}^\prime }^{\phi,JIY}
= \sum_{\mu,\alpha} 
\left( \begin{array}{cc|c} \treintaycinco& \cincuentayseis&
   \phi \\ \mu_M J_M & \mu_B J_B &
   \mu J \alpha \end{array}\right)\nonumber\\
&\times& 
\left( \begin{array}{cc|c}
 \mu_M& \mu_B& \mu \\ 
  I_M Y_M & I_B Y_B & I Y\end{array}\right)
\left( \begin{array}{cc|c} \mu^\prime_{M^\prime}& \mu_{B^\prime}^\prime&
   \mu \\ I^\prime_{M^\prime}Y^\prime_{M^\prime} &
 I^\prime_{B^\prime} Y^\prime_{B^\prime} &
   IY\end{array}\right) 
\nonumber\\
&\times&
\left( \begin{array}{cc|c} \treintaycinco& \cincuentayseis&
   \phi \\ \mu'_{M'} J'_{M'} & \mu'_{B'} J'_{B'} &
   \mu J \alpha \end{array}\right).
 \phantom{hhhhhhhhhhhhh}
\end{eqnarray}
\section{Chiral Symmetry Constraints}

We make use of the underlying CS and propose a chiral expansion to
determine the $V_{\phi}(s)$ functions. Thus, we look at
the effective $s$-wave {\it potential} describing the interaction of
the Goldstone pseudoscalar meson and the lowest $J^P=\frac12^+$ baryon
octets. From the SU(3) WT chiral Lagrangian (we use the convention $V=-{\cal
L}$), one finds for each $(I,Y)$ sector [(0,2), (1,2), (1/2,1),
(3/2,1), (0,0),(1,0),(2,0),(1/2,$-$1),(3/2,$-$1),(0,$-$2),(1,$-$2)] and on the
mass shell (recall that $J=1/2$)
\begin{eqnarray}
V^{IY}_{ab}(\sqrt{s})& = &
D^{IY}_{ab} \frac{2\,\sqrt{s}-M_a-M_b}{4\,f^2} 
\label{eq:lowest}
\end{eqnarray}
with
\begin{eqnarray}
D^{IY} &=& \sum_{\mu,\gamma,\gamma^\prime}
\lambda_{\mu_\gamma \to \mu_{\gamma^\prime}} \left (
\begin{array}{cc|c} \ocho & \ocho & \mu_\gamma \\ 
I_M Y_M & I_B Y_B & IY
\end{array}
\right)
\nonumber \\
&\times& \left (
\begin{array}{cc|c} \ocho & \ocho & \mu_{\gamma^\prime} \\ 
I^\prime_{M^\prime} Y^\prime_{M^\prime} & I^\prime_{B^\prime} Y^\prime_{B^\prime} &
  IY\end{array}\right) 
\,,
\end{eqnarray}
where $M_b$ ($M_a$) is the baryon mass of the initial (final) channel,
$f\simeq 93\,$MeV the pion weak decay constant, $\mu$ runs over the
$\veintisiete$, $\diez$, $\diez^*$, $\ocho$ and $\uno$ SU(3) representations and
$\gamma,\gamma^\prime$ are used to account for the two octets ($\ocho_s$
and $\ocho_a$) which appear in the $\ocho\otimes \ocho$ decomposition
($\veintisiete+\diez+\diez^*+\ocho_s+\ocho_a+\uno$).
Besides, $\lambda_{\veintisiete}= 2$,
$\lambda_{\ocho_s}= \lambda_{\ocho_a}=-3$, $\lambda_{\uno}=-6$, $\lambda_{\diez} =
\lambda_{\diez^*}=\lambda_{\ocho_s\leftrightarrow \ocho_a} =0$, which reproduces
the $D$-matrix eigenvalues found in Ref.~\cite{Garcia-Recio:2003ks}. Thus, we see that
CS at leading order is much more predictive than SU(3) symmetry, and
it predicts the values of the seven $\lambda$ couplings, which
otherwise will be totally arbitrary functions of $s$. Note that the
SU(3) WT Lagrangian also provides the $s$ dependence ($(\sqrt{s}-M)$, with
$M$ the common mass of the baryon octet in the SU(3) limit) of the
effective potential, and thus one is left with only two free
parameters, namely, $f$ and $M$.

It is clear that not all SU(3) invariant interactions in the
$(\ocho_\uno)$meson--$(\ocho_\dos)$baryon sector can be extended to a SU(6)
invariant interaction. Remarkably, the seven couplings ($\lambda$'s)
in the WT interaction turn out to be consistent with SU(6) and
moreover, the extension is unique. In other words, there is a choice
of the {\em four} couplings for the $\treintaycinco \otimes
\cincuentayseis$ interaction that, when restricted to the $\ocho_\uno \otimes
\ocho_\dos$ sector, reproduces the {\em seven} SU(3) WT couplings and such choice
is unique. Indeed, the {\it potential} of Eq.~(\ref{eq:lowest}) can be
recovered, in the SU(3) limit, from Eq.~(\ref{eq:su6}) by taking
\begin{equation}
V_{\phi}(s) = \bar\lambda_{\phi}
\frac{\sqrt{s}-M}{2\,f^2}\,,
\label{eq:vsu6} 
\end{equation}
with $\bar\lambda_{\cincuentayseis}=-12$,
$\bar\lambda_{\setenta}=-18$, $\bar\lambda_{\setecientos}=6$ and
$\bar\lambda_{\milcientotreintaycuatro}=-2$ and $M$ now being the
common octet and decuplet baryon mass.\footnote{The SU(6) extension
thus obtained (Eqs.~(\ref{eq:su6}) and~(\ref{eq:vsu6})) also leads to
the {\it potentials} used in Ref.~\cite{Kolomeitsev:2003kt,Sarkar:2004jh} to study the
($\diez_\cuatro$)baryon--($\ocho_\uno$)meson sector. Note also that
the $\setenta$ of SU(6) leads to the most attractive $s$-wave
meson--baryon interaction. This would be consistent with the scenario
studied in~\cite{Goity:2002pu}, namely, that the first negative parity baryon
excited states are members of the $\setenta$ multiplet, and also with 
constituent quark model considerations, where the $\setenta$-plet appears
when one of the quarks occupies  an orbitally excited level \cite{Gibson:1976bk}.\label{foot:1}}
The underlying reason for this is CS. Indeed, the WT Lagrangian is not
just SU(3) symmetric but also chiral
($\text{SU}_L(3)\otimes\text{SU}_R(3)$) invariant. Symbolically (and
up to an overall coefficient)
\begin{equation}
{\cal L_{\rm WT}}= {\rm Tr} ( [M^\dagger, M]B^\dagger B)
\,.
\end{equation}
This structure, dictated by CS, is more suitably analyzed
in the $t$-channel. The mesons $M$ fall in the representation $\ocho$
which is also the adjoint representation. The commutator
$[M^\dagger,M]$ indicates a $t$-channel coupling to the $\ocho_a$
(antisymmetric) representation, thus
\begin{equation}
{\cal L_{\rm WT}}= \left((M^\dagger\otimes M)_{\ocho_a}\otimes
(B^\dagger\otimes B)_{\ocho}\right)_{\uno}
\,.
\end{equation}
The unique SU(6) extension is then
\begin{equation}
{\cal L_{\rm WT, SU(6)}}= \left((M^\dagger\otimes M)_{\treintaycinco_a}\otimes
 (B^\dagger\otimes B)_{\treintaycinco}\right)_{\uno}
\,,
\label{eq:1}
\end{equation}
since the $\treintaycinco$ is the adjoint representation of SU(6). The $t$-channel
decompositions $\treintaycinco \otimes \treintaycinco = 
\uno \oplus \treintaycinco_s\oplus \treintaycinco_a\oplus \cientoochentaynueve
\oplus \doscientosochenta \oplus \doscientosochenta^* \oplus \cuatrocientoscinco$ and 
$\cincuentayseis \otimes \cincuentayseis^* 
= \uno \oplus \treintaycinco \oplus \cuatrocientoscinco \oplus \dosmilseiscientosnoveintaycinco$
indicate that the coupling in Eq.~(\ref{eq:1})
exists and is indeed unique, all coupling constants being reduced to a
single independent one, namely, that of the WT Lagrangian (pion weak
decay constant, besides the hadron masses).
\begin{figure}[b]
\begin{center}
\makebox[0pt]{\input{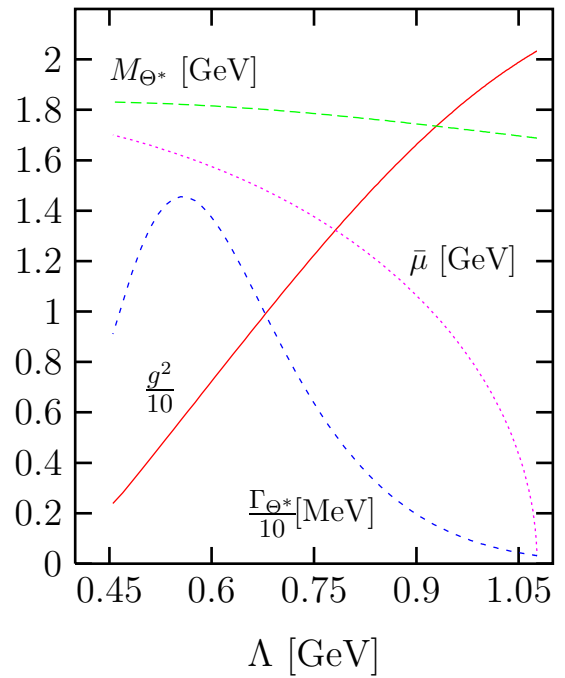}}
\end{center}
\vspace{-0.5cm}
\caption{Resonance $\Theta^{*+}$ properties as a function of the
  UV cutoff $\Lambda $ or the subtraction scale ${\bar \mu}$.}
\label{fig:fig1}
\end{figure}

The large $N_c$ behavior of the SU(6) WT extension proposed here has been
contemplated in \cite{Garcia-Recio:2006wb}. Two interesting
conclusions of that study are, i) a consistent treatment of the WT
interaction yields a generic large $N_c$ amplitude of the same order
as the baryon pole one, ${\cal O}(N_c^0)$, instead of a $1/N_c$
suppression, as usually assumed. And ii) the SU(6) WT interaction is large
in the $\setenta$-plet SU(6) sector even for large $N_c$, which, besides
being consistent with quark model considerations \cite{Gibson:1976bk} (see also
footnote \ref{foot:1}), solves the conflict between the
phenomenological success of BSE approaches using the WT mechanism
\cite{Kaiser:1995eg,Oset:1997it,Lutz:2003fm,Jido:2003cb,Nieves:2001wt,Garcia-Recio:2002td,%
Garcia-Recio:2003ks,Kolomeitsev:2003kt,Sarkar:2004jh} and the previously assumed
suppression of this interaction in the large $N_c$ limit.

\section{Meson--Baryon Scattering Matrix and the $\Theta^{*+}$ Baryon.}

We solve the coupled channel BSE with an interaction kernel determined
by Eqs.~(\ref{eq:su6}) and~(\ref{eq:vsu6}). Some mass breaking effects
can be taken into account just by replacing $(\sqrt{s}-M)$ by
$(2\sqrt{s}-M_a-M_b)/2$ in Eq.~(\ref{eq:vsu6}). In a given $JIY$
sector, the solution for the coupled channel $s$-wave scattering
amplitude, $T^{J}_{IY}(\sqrt{s})$ (normalized as the $t$ matrix
defined in Eq.~(33) of the first entry of Ref~\protect\cite{Oset:1997it,Lutz:2003fm,Jido:2003cb}),
in the {\it on-shell} scheme~\cite{Nieves:1998hp,Nieves:1999bx,%
Oset:1997it,Lutz:2003fm,Jido:2003cb,Oset:1997it,Lutz:2003fm,Jido:2003cb}
reads,
\begin{eqnarray}
T^J_{IY}(\sqrt{s}) &=& \frac{1}{1-
V^J_{IY}(\sqrt{s})\,J^J_{IY}(\sqrt{s})}\,V^J_{IY}(\sqrt{s})
 \label{eq:scat-eq}
\end{eqnarray}
with
\begin{eqnarray}
V^J_{IY}(\sqrt{s}) &=& \langle {\cal M}^\prime  {\cal B}^\prime ; JIY
| V| {\cal M} {\cal B} ; JIY \rangle
\,,
\end{eqnarray}
and $J^J_{IY}(\sqrt{s})$ a diagonal matrix of loop
functions~\protect\cite{Nieves:2001wt,Garcia-Recio:2002td}, which logarithmically diverge and
hence need one subtraction or an ultraviolet (UV) cutoff to make them
finite.  Possible $d$-wave mixings (chiral corrections 
provide $d$-wave meson--baryon amplitudes) and the rich new
phenomenology (there are new open channels, due to the inclusion of
the vector meson degrees of freedom and the products from their
decays, not taken into account in the usual meson--baryon analysis
based on the WT interaction) which can be extracted from
Eq.~(\ref{eq:scat-eq}) will be studied elsewhere.  We will focus here
on the $Y=2$ (strangeness=+1) sector of great interest nowadays, since
the claim by the LEPS collaboration of the observation of the
$\Theta^+(1540)$ resonance. Though its existence is still under
discussion and it needs to be confirmed, it seems clear that the
possible candidate would be an isoscalar, extremely narrow (with width
definitely smaller than $10\,$MeV), while its spin-parity has not been
established yet~\cite{Nakano:2003qx}.

In the $Y=2$, $I=0$ sector, the $|K N\rangle $ and $|K^*
N\rangle $ states appear. We have $|K N\rangle_{J=1/2}=\left ( |\setecientos;
\diez^*_\dos\rangle 
- \sqrt{3}\,|\milcientotreintaycuatro; 
\diez^*_\dos\rangle \right)/2 $ 
and $|K^* N\rangle_{J=1/2}=\left (-\sqrt{3}\, |\setecientos; \diez^*_\dos\rangle 
- |\milcientotreintaycuatro;
\diez^*_\dos\rangle \right)/2 $, 
and $|K^* N\rangle_{J=3/2}= - |\milcientotreintaycuatro; \diez^*_\cuatro
\rangle$. 
In the $J=1/2$
channel, we find a resonance (pole in the second Riemann
sheet~\cite{Nieves:2001wt,Garcia-Recio:2002td}), though its exact position depends on the details
of the Renormalization Scheme (RS) used. It comes out too wide
($\Gamma > 100\,$MeV) to be identified as the $\Theta^+(1540)$. The
situation is much more suggestive for $J=3/2$. There, the effective
interaction is determined by the $\milcientotreintaycuatro$ SU(6) representation alone and
it is attractive (${\bar\lambda}_{\milcientotreintaycuatro}=-2$). A pole is found in the
first Riemann sheet corresponding to a $K^* N$ bound state which we
call $\Theta^{*+}$. This state is unstable since the $K^*$ decays into
$K\pi$. In order to estimate the $\Theta^{*+}$ width, 
we model the $\Theta^{*}N K^*$ coupling as
\begin{eqnarray}
{\cal L}_{\Theta^* N K^*}&=& 
-\frac{g}{\sqrt{2}} 
\overline{\Theta}^\mu 
 \left(  K^{*0}_\mu p - K^{*+}_\mu n \right) 
 + \text{h.c.},
\end{eqnarray}
while the $K^*$ decay is described by
\begin{eqnarray}
{\cal L}_{K^* K \pi}&=& 
-\frac{i{g^\prime}}{\sqrt 2 f^2}\Big \{ 
\partial^\mu\pi^+\partial^\nu K^- 
\left(\partial_\mu K^{*0}_\nu-\partial_\nu K^{*0}_\mu  \right)
\nonumber \\
&+&  
\partial^\mu\pi^-\partial^\nu {\bar K}^0
\left(\partial_\mu K^{*+}_\nu-\partial_\nu K^{*+}_\mu  \right)
\nonumber \\ 
&-&\frac{1}{\sqrt 2}
\partial^\mu\pi^0\partial^\nu {\bar K}^0
\left ( \partial_\mu K^{*0}_\nu - \partial_\nu K^{*0}_\mu \right)
\\ 
&+& \frac{1 }{\sqrt 2}
\partial^\mu\pi^0\partial^\nu K^-
\left( \partial_\mu K^{*+}_\nu - \partial_\nu K^{*+}_\mu \right)
 \Big\}  + \text{h.c.}
\nonumber 
\end{eqnarray}
In the previous formulas, $\Theta^\mu$ is a Rarita-Schwinger field,
$p$ and $n$ the nucleon fields, $ K^{*0}_\mu$, $ {\bar K}^{*0}_\mu =
(K^{*0}_\mu)^\dagger $ and $K^{*+}_\mu = (K^{*-}_\mu)^\dagger$ the
Proca fields which annhilate and create neutral and charged $K^*$ and
${\bar K}^*$ mesons, and similarly for the kaon and antikaon
fields, while $\pi^{0}$ and $\pi^+=(\pi^-)^\dagger$ are
the pion fields. 
The coupling $g$ is determined by the residue at the pole of
$T^{\frac32}_{02}$ [i.e., $T^{\frac32}_{02} \approx g^2 \times
2M_{\Theta^*}/(s-M^2_{\Theta^*})$] and we fix ${g^\prime}\approx 0.14$
to reproduce $\Gamma_{K^{*0}}\approx\Gamma_{K^{*+}}\approx 50\,$MeV
(we use charged averaged masses). This gives
\begin{eqnarray}
\Gamma_{\Theta^*}&=&
\frac{g^2{g^\prime}^2}{32\pi^3 f^4}\frac{1}{6M_{\Theta^*}}
\int_{m_\pi+m_K}^{M_{\Theta^*}-M_N} 
\!\! d{\tilde m}\, {\tilde m}^2\,{\tilde q}\, q_\pi^3
\nonumber \\  && \times
\frac{(2 {\tilde m}^2+{\tilde E}^2) (M_{\Theta^*}+M_N-{\tilde E})
}{ 
({\tilde m}^2-m^2_{K^*})^2 + ({\tilde m}{\tilde \Gamma})^2
}
\,,
\end{eqnarray}
where ${\tilde m}$, ${\tilde E}$, ${\tilde q}$, and ${\tilde \Gamma}$
are the  invariant mass, energy, momentum, and width,
respectively, of the virtual $K^*$ in the at rest $\Theta^*$ system,
\begin{eqnarray}
\tilde E &=&  \frac{M_{\Theta^*}^2+{\tilde m}^2-M_N^2}{2M_{\Theta^*}} \,,
\nonumber \\
{\tilde q}^2 &=&  {\tilde E}^2 - {\tilde m}^2 \,,
\nonumber \\
{\tilde \Gamma} &=&  \frac{{g^\prime}^2}{32\pi f^4} q_\pi^3 {\tilde m}^2  \,,
\end{eqnarray}
and $q_\pi$ is the pion momentum in the at rest $K^*$ system
\begin{equation}
q_\pi^2 = \frac{\lambda({\tilde m}^2,m_\pi^2,m^2_K)}{(2{\tilde m})^2}  \,.
\end{equation}
Resonance mass, residue and width depend on the RS employed. We have
used an UV cutoff ($\Lambda$) to evaluate the loop function
$J(\sqrt{s})$, which is equivalent to choose an scale ${\bar \mu}$
such that $J(\sqrt{s}={\bar \mu})=0$. Results are shown in
Fig.~\ref{fig:fig1}. For ${\bar \mu}$ ranging from $0.05\,$GeV
($\Lambda\approx 1.08\,$GeV) to $1.7\,$GeV ($\Lambda\approx
0.46\,$GeV) the resonance mass (width) varies from $1.688\,$GeV
($0.3\,$MeV), close to the $(M_N+m_\pi+m_K)$ threshold, to
$1.831\,$GeV ($9\,$MeV, but the width does not grow monotonously, see
figure), close to the $(M_N+m_{K^*})$ threshold.

In our treatment we have not included $d$-wave ${K} N$, ${K}^* N$
contributions, nor further $s$-wave terms such as the $u$-channel pole
graph or single pion exchange between $K^*$ and $N$ (the vertex
$K^*K^*\pi$ being of abnormal parity) since none of these mechanisms
contributes to the $K^* N$ $s$-wave scattering length (i.e., they
vanish at threshold)\footnote{Except for a small contribution from the
antibaryon term of the pole graph, all elementary vertices involved
are $p$-wave.}. Another mechanism for $K^*N$ scattering not included
here would be the sequential exchange of two pions with an
intermediate $K$ meson, corresponding to a box graph $K^*N\to KN \to
K^*N$, which involves two $p$-wave normal parity $K^*K\pi$
vertices. Unlike the single pion exchange, such a contribution is not
vanishing at threshold since the two virtual pions need not carry a
small momentum. An analogous mechanism has been considered long ago
for $KN$ scattering \cite{Aaron:1971ka,Alcock:1977ep}, this time with
the box graph $KN\to K^*N \to KN$. Technically a box graph is
difficult to work with, since one must somehow renormalize its
intrinsic UV divergence, and then renormalize its contribution in the
BSE ladder. This is quite hard for a non contact-like vertex and, at
present, certainly beyond a consistent chiral unitary
treatment. Fortunately, available calculations of $\bar{K}N$
scattering within the BSE chiral unitary approach (none of them
including box graphs \cite{Oset:1997it,Lutz:2003fm,Jido:2003cb,Nieves:2001wt,%
Garcia-Recio:2002td,Garcia-Recio:2003ks,Kolomeitsev:2003kt,Sarkar:2004jh}) yield a fairly good
quantitative description of $s$-wave baryonic resonance data. This
would suggest that the box mechanism would not play a crucial role. In
summary, we do not expect the corrections to the mass and width
estimated above for the $\Theta^*$ resonance to be large enough to
affect its existence.
Possible production and identification mechanisms for this resonance
could be found in reactions like $\gamma p \to {\bar K}^0 p K^+ \pi^-$
by meassuring the three body $p K^+ \pi^-$ invariant mass.

The scheme presented here also contains other exotic states which will
be examined elsewhere. For instance in the $Y=-3$, $I=J=1/2$ sector we
find 
$|{\bar K}^* \Omega \rangle= |\milcientotreintaycuatro; \treintaycinco_\dos \rangle$
and thus we
have an attractive ${\bar K}^* \Omega$ interaction and possibly a
bound state.

\begin{acknowledgments}
{\footnotesize
This work was supported by DGI, FEDER, UE and Junta de Andaluc{\'\i}a funds 
(FIS2005-00810, HPRN-CT-2002-00311, FQM225).}
\end{acknowledgments}


\end{document}